\documentclass[aps,prl,twocolumn,superscriptaddress]{revtex4-2}

\usepackage{graphicx}
\usepackage{booktabs}

\begin{document}


\title{Neutron transfer reactions on the ground state and isomeric state of a $^{130}$Sn beam}


\author{K.L.~Jones}
\email{kgrzywac@utk.edu}
\affiliation{Department of Physics and Astronomy, 401 Nielsen Physics Building, 1408 Circle Drive, University of Tennessee, Knoxville, TN 37996, USA }
\author{A.~Bey}
\affiliation{Department of Physics and Astronomy, 401 Nielsen Physics Building, 1408 Circle Drive, University of Tennessee, Knoxville, TN 37996, USA }
\affiliation{Joint Institute for Nuclear Physics and Applications, Oak Ridge National Laboratory, Oak Ridge TN 37831, USA}
\author{S.~Burcher}
\affiliation{Department of Physics and Astronomy, 401 Nielsen Physics Building, 1408 Circle Drive, University of Tennessee, Knoxville, TN 37996, USA }
\author{J.M.~Allmond}
\affiliation{Joint Institute for Nuclear Physics and Applications, Oak Ridge National Laboratory, Oak Ridge TN 37831, USA}
\affiliation{Physics Division, Oak Ridge National Laboratory, Oak Ridge, TN 37831, USA}
\author{A. Galindo-Uribarri}
\affiliation{Physics Division, Oak Ridge National Laboratory, Oak Ridge, TN 37831, USA}
\affiliation{Department of Physics and Astronomy, 401 Nielsen Physics Building, 1408 Circle Drive, University of Tennessee, Knoxville, TN 37996, USA }
\author{D.C. Radford}
\affiliation{Physics Division, Oak Ridge National Laboratory, Oak Ridge, TN 37831, USA}
\author{S.~Ahn}
\affiliation{Department of Physics and Astronomy, 401 Nielsen Physics Building, 1408 Circle Drive, University of Tennessee, Knoxville, TN 37996, USA }
\affiliation{Joint Institute for Nuclear Physics and Applications, Oak Ridge National Laboratory, Oak Ridge TN 37831, USA}
\author{A.~Ayres}
\affiliation{Department of Physics and Astronomy, 401 Nielsen Physics Building, 1408 Circle Drive, University of Tennessee, Knoxville, TN 37996, USA }
\author{D.W.~Bardayan}
\affiliation{Physics Division, Oak Ridge National Laboratory, Oak Ridge, TN 37831, USA}
\affiliation{Department of Physics, University of Notre Dame, Notre Dame, IN 46556, USA}
\author{J. A.~Cizewski}
\affiliation{Department of Physics and Astronomy, Rutgers University, New Brunswick, NJ 08903, USA}
\author{R.F.~Garcia Ruiz}
\altaffiliation{Current address Massachusetts Institute of Technology, Cambridge, MA 02139, USA}
\affiliation{Joint Institute for Nuclear Physics and Applications, Oak Ridge National Laboratory, Oak Ridge TN 37831, USA}
\affiliation{Physics Division, Oak Ridge National Laboratory, Oak Ridge, TN 37831, USA}
\affiliation{Instituut voor Kernen Stralingsfysica, KU Leuven, B-3001, Leuven, Belgium}
\author{M.E.~Howard}
\affiliation{Department of Physics and Astronomy, Rutgers University, New Brunswick, NJ 08903, USA}
\author{R.L.~Kozub}
\affiliation{Department of Physics, Tennessee Technological University, Cookeville, TN 38505, USA}
\author{J.F.~Liang}
\affiliation{Physics Division, Oak Ridge National Laboratory, Oak Ridge, TN 37831, USA}
\author{B.~Manning}
\affiliation{Department of Physics and Astronomy, Rutgers University, New Brunswick, NJ 08903, USA}
\author{M.~Matos}
\affiliation{Department of Physics and Astronomy, Louisiana State University, Baton Rouge, LA 70803, USA}
\author{C.D.~Nesaraja}
\affiliation{Physics Division, Oak Ridge National Laboratory, Oak Ridge, TN 37831, USA}
\author{P.D.~O'Malley}
\affiliation{Department of Physics, University of Notre Dame, Notre Dame, IN 46556, USA}
\affiliation{Department of Physics and Astronomy, Rutgers University, New Brunswick, NJ 08903, USA}
\author{E.~Padilla-Rodal}
\affiliation{Instituto do Ciencias Nucleares, UNAM, AP 70-543, 04510, Mexico}
\author{S.D.~Pain}
\affiliation{Physics Division, Oak Ridge National Laboratory, Oak Ridge, TN 37831, USA}
\author{S.T.~Pittman}
\affiliation{Physics Division, Oak Ridge National Laboratory, Oak Ridge, TN 37831, USA}
\affiliation{Department of Physics and Astronomy, Louisiana State University, Baton Rouge, LA 70803, USA}
\author{A.~Ratkiewicz}
\affiliation{Department of Physics and Astronomy, Rutgers University, New Brunswick, NJ 08903, USA}
\author{K.T.~Schmitt}
\affiliation{Department of Physics and Astronomy, 401 Nielsen Physics Building, 1408 Circle Drive, University of Tennessee, Knoxville, TN 37996, USA }
\author{M.S.~Smith}
\affiliation{Physics Division, Oak Ridge National Laboratory, Oak Ridge, TN 37831, USA}
\author{D.W.~Stracener}
\affiliation{Physics Division, Oak Ridge National Laboratory, Oak Ridge, TN 37831, USA}
\author{R.L.~Varner}
\affiliation{Physics Division, Oak Ridge National Laboratory, Oak Ridge, TN 37831, USA}



\date{\today}

\begin{abstract}
 The structure of nuclei around the neutron-rich nucleus $^{132}$Sn is of particular interest due to the vicinity of the Z~=~50 and N~=~82 shell closures and the r-process nucleosynthetic path.  Four states in $^{131}$Sn with a strong single-particle-like component have previously been studied via the (d,p) reaction, with limited excitation energy resolution.  The $^{130}$Sn($^{9}$Be,$^{8}$Be)$^{131}$Sn and $^{130}$Sn($^{13}$C,$^{12}$C)$^{131}$Sn single-neutron transfer reactions were performed in inverse kinematics at the Holifield Radioactive Ion Beam Facility using particle-$\gamma$ coincidence spectroscopy. The uncertainties in the energies of the single-particle-like states have been reduced by more than an order of magnitude using the energies of $\gamma$ rays.  The previous tentative J$^{\pi}$ values have been confirmed.  Decays from high-spin states in $^{131}$Sn have been observed following transfer on the isomeric component of the $^{130}$Sn beam.  The improved energies and confirmed spin-parities of the p-wave states important to the r-process lead to direct-semidirect cross-sections for neutron capture on the ground state of $^{130}$Sn at 30~keV that are in agreement with previous analyses.  A similar assessment of the impact of neutron-transfer on the isomer would require significant nuclear structure and reaction theory input.  There are few measurements of transfer reaction on isomers, and this is the first on an isomer in the $^{132}$Sn region.
\end{abstract}


\maketitle


\section{Introduction}
Knowledge of the structure of exotic tin nuclei near the neutron shell closures is important for characterizing nuclear models away from the valley of stability, such as the nuclear shell model \cite{May55}, and modern methods including In-Medium Similarity Renormalization (IMSGR) \cite{Miy21} and coupled cluster \cite{Mor18}. Information about the nature of single-particle and single-hole states outside the doubly-magic core of $^{132}$Sn, for example, is essential to predictions of many nuclei that are not currently available for measurement.  This includes having accurate excitation energies and J$^{\pi}$ assignments in order to test model predictions of spectra.  Transfer reactions are a powerful tool for measuring the single-particle-like structure of short-lived nuclei.

The structure around $^{132}$Sn is also important for understanding the astrophysical rapid neutron capture ({\em r-}) process, which is responsible for the production of about half the elements heavier than Fe \cite{Bur57}. Following recent evidence for neutron star mergers as a site of the r-process \cite{Tan17}, the uncertainties in element production from this scenario are weighted toward nuclear data.  
However, recent evidence of two populations of stars with different r-process abundances \cite{Tsu17} suggests that there is not a unique site of the r-process.  
Core-collapse supernovae may be needed to explain r-process elements in low-metallicity environments, such as were present in the early Universe  \cite{Hal18}.   
In both neutron star merger and core-collapse supernova scenario, the abundances of heavy elements are sensitive to $\beta$-decay half lives and neutron-capture rates of certain bottleneck nuclei that have long half-lives and inhibit the flow of material back to stability via $\beta$ decay.   

\begin{figure}[ht]
       \includegraphics[width=0.95\linewidth]{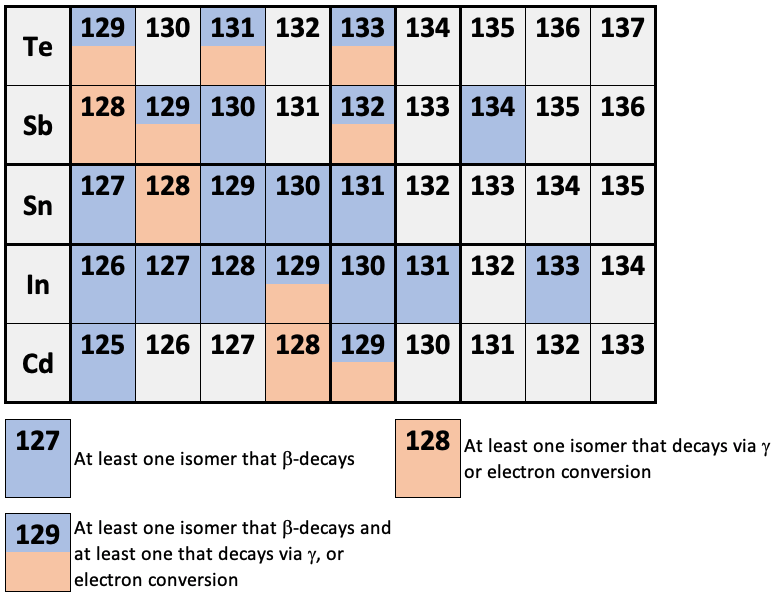}
        \caption{Schematic of the portion of the chart of the nuclides from $^{125}$Cd to $^{137}$Te.  The color coding indicates the presence of at least one known isomer that has a significant $\beta$ branch (blue), or decays via either $\gamma$, or conversion-electron and has a t$_{1/2}>$ 1~ms (coral)\cite{NNDC} (color online). }
        \label{fig:isomers}
\end{figure}

The neutron-capture cross section of the bottleneck nucleus $^{130}$Sn ($t_{1/2} = 3.72$~minutes) has been shown to have a large influence on the shape of the r-process abundance pattern \cite{Beu09, Sur09}.  
An added uncertainty in nucleosynthesis calculations comes from astromers - astrophysically important nuclear isomeric states \cite{Apr05, Mis20}. 
For an isomer to be an astromer it has to behave in a significantly different way to the ground state and either survive long enough to undergo a reaction, or undergo $\beta$ decay at a different rate to the ground state.  
Theorists have started working on identifying and treating astromers in the r-process as well as other nucleosynthetic processes \cite{Rei18, Fuj20}.  
With the occurrence of many low-lying isomers around $^{132}$Sn, as highlighted in  \cite{Fuj20} and \cite{Mar99}, and displayed in Figure \ref{fig:isomers}, the influence of these long-lived states could be profound.  
The isomeric 7$^-$ state at $E_{x}=1.95$~MeV in $^{130}$Sn has a t$_{1/2}$ = 1.7 minutes, long enough to be considered stable on an r-process time scale.

Measuring neutron-capture cross sections on the ground or isomeric states of fission fragments would require a target of either the short-lived isotope or of neutrons ($\tau$ = 879.6 $\pm$ 0.8 s \cite{PDG}). 
For this reason, the vast majority of the neutron-capture rates important to the r-process are calculated from masses and J$^{\pi}$ assignments from either measurements or theoretical models.  
Many of the nuclei on the r-process path are outside the reach of current facilities, and although the mass and decay properties of many nuclei can be measured in a single campaign \cite{Lor15}, it will be many years before experimental spectroscopic information on these very exotic nuclei will be available. \\ 
Direct capture (DC) is expected to play an important role in neutron capture near $^{132}$Sn.  
DC depends sensitively on the energies and spectroscopic strengths of the p$_{3/2}$ and p$_{1/2}$ states.  These properties are informed by the present work that constrains the excitation energies and confirms the J$^{\pi}$ assignments of the 3/2$^-$ and 1/2$^-$ states and previous measurements near $^{132}$Sn \cite{Jon11, Koz12, Man19}\nocite{Man19err}.

A measurement of the $^{130}$Sn(d,p)$^{131}$Sn reaction in inverse kinematics \cite{Koz12} identified four states with properties similar to the low-lying neutron single-particle states in $^{133}$Sn \cite{Hof96,Urb99,Jon10}, built upon two holes below the $N=82$ shell closure.  These states in $^{131}$Sn were tentatively assigned  to the 2$f_{7/2}$, 3$p_{3/2}$, 3$p_{1/2}$, and 2$f_{5/2}$ single-particle states coupled to two holes below the $N=82$ shell closure.  The spin-parity assignments were based on the measured proton angular distributions, which give the transferred ${\ell}$-values, and comparison to the single-particle states from the normal shell model ordering.  There have been two recent studies of states in $^{131}$Sn produced by removing a single neutron from the $N=8$2 core via the $^{132}$Sn(d,t)$^{131}$Sn reaction \cite{Orl18} and the one-neutron knockout reaction \cite{Vaq20}.  These removal mechanisms do not favor the population of the single-particle-like states in $^{131}$Sn that are the subject of the current study.

Here, measurements of $\gamma$ rays following the single-neutron-adding ($^9$Be,$^8$Be) and ($^{13}$C,$^{12}$C) reactions on a beam of $^{130}$Sn are presented.  By measuring $\gamma$ rays, we are able to provide more precise energies for the states that decay to the ground state and precise energies above the 11/2$^-$ isomer for the states that decay to it.  Additionally, the comparison of the relative intensities of $\gamma$ rays depopulating the same states in the beryllium and carbon target measurements is used to constrain the J$^{\pi}$ assignments of those states.
\section{Experimental procedure}
The experiment conducted at the Holifield Radioactive Ion Beam Facility (HRIBF) \cite{Bee11} at Oak Ridge National Laboratory used a beam of $^{130}$Sn from proton-induced fission on a Uranium Carbide target.  A surface ionization source containing H$_{2}$S gas produced SnS$^{+}$ molecules, which were accelerated in the 25-MV electrostatic tandem.  The first-stage separator was tuned for mass = 162, significantly reducing the mass = 130 isobars that do not readily form sulfide molecules.  After charge exchange and breakup of the SnS molecules, the $^{130}$Sn ions were accelerated to a total energy of 520~MeV. The beam was $\approx$92\% pure $^{130}$Sn with small components of the isobaric contaminants $^{130}$Te, $^{130}$Sb, and $^{130}$I. Previous measurements of the isomeric content of $^{130}$Sn beams showed a range of  8.7\%  to 13.1\% of the $^{130}$Sn beam was delivered in the 7$^{-}$ isomeric state \cite{Str} depending on the type of beam accelerated e.g. atomic, molecular.  Specifically, 9.6\% of negative $^{130}$Sn ions from the breakup of SnS in a previous measurement were found to be in the isomeric state.  The beam, with an average intensity of $10^5$ pps impinged on targets of 2-mg/cm$^2$ $^{nat}$Be and 2-mg/cm$^2$ $^{13}$C for 45 and 52 hours, respectively.

The $\gamma$ decays were measured in the CLARION array  of 11 Compton-suppressed HPGe clover detectors positioned at 90$^{\circ}$, 132$^{\circ}$, and 154$^{\circ}$ in the laboratory frame \cite{Gro00} with a total efficiency of 3.00(5)\% at 1~MeV.  The efficiency of the array was found using calibrated sources and characterized using a simulation of CLARION. Recoiling target-like nuclei and scattered beam particles were detected in the BareBall CsI(Tl) array \cite{Gal10} of four concentric rings covering angles from 7$^{\circ}$ to 60$^{\circ}$.  Pulse-shape discrimination was used to produce particle identification plots to differentiate reaction products.
  In the case of the $^{9}$Be target, the unbound $^{8}$Be ejectile promptly breaks up into two $\alpha$ particles, with a relatively small decay energy of 92~keV.   The two $\alpha$ particles could either hit the same segment of BareBall, leading to the 2$\alpha$ gate, or they could hit two adjacent segments, resulting in two hits in the $\alpha$ gate.  Both situations provide a clean tag of the reaction channel involving a $^8$Be ejectile.  The $^{12}$C ejectiles can be partially separated from the $^{13}$C target nuclei although it is less clean than the $^{9}$Be case.  
The Doppler correction was made according to the ring of BareBall where the emergent $\alpha$ particles, or $^{12}$C nucleus, were detected, which relates directly to the angle and nominal $\beta =$v/c of the recoiling $^{131}$Sn.  The spectrum of $\gamma$ rays measured in coincidence with either the 2~$\alpha$ gate, or two $\alpha$ particles in neighboring detectors of BareBall, following the $^{130}$Sn($^{9}$Be, $^{8}$Be $\gamma$)$^{131}$Sn reaction is shown in Figure \ref{fig:spectrum}.  
As the recoils were $\gamma$ decaying in-flight, the $\beta$ at the time of emission depends on the lifetime of the state.  The level of statistics was not high enough to extract precise lifetimes as was achieved  for $^{133}$Sn with the same method \cite{All14}.  Instead, the decays were determined to be either ``slow" or ``fast" decays depending on the width of the line shapes using two Doppler corrections. The ``slow" Doppler correction assumes that the state decays after significant energy loss in the target and was taken from the experimental $\beta$ for the 331.7~keV state. The ``fast" correction assumes that the state decays immediately without significant slowing of the recoil and was taken from the 1895-keV transition. The lowest energy $\gamma$ rays are shown in panel (a) of Fig. \ref{fig:spectrum} using the ``slow" correction.  Panels (b) and (c) are corrected using the ``fast" Doppler correction.  

\section{Experimental Results}
\begin{figure}[ht]
   \includegraphics[trim=0cm 0cm 0cm 0.2cm, width=1.05\linewidth,clip]{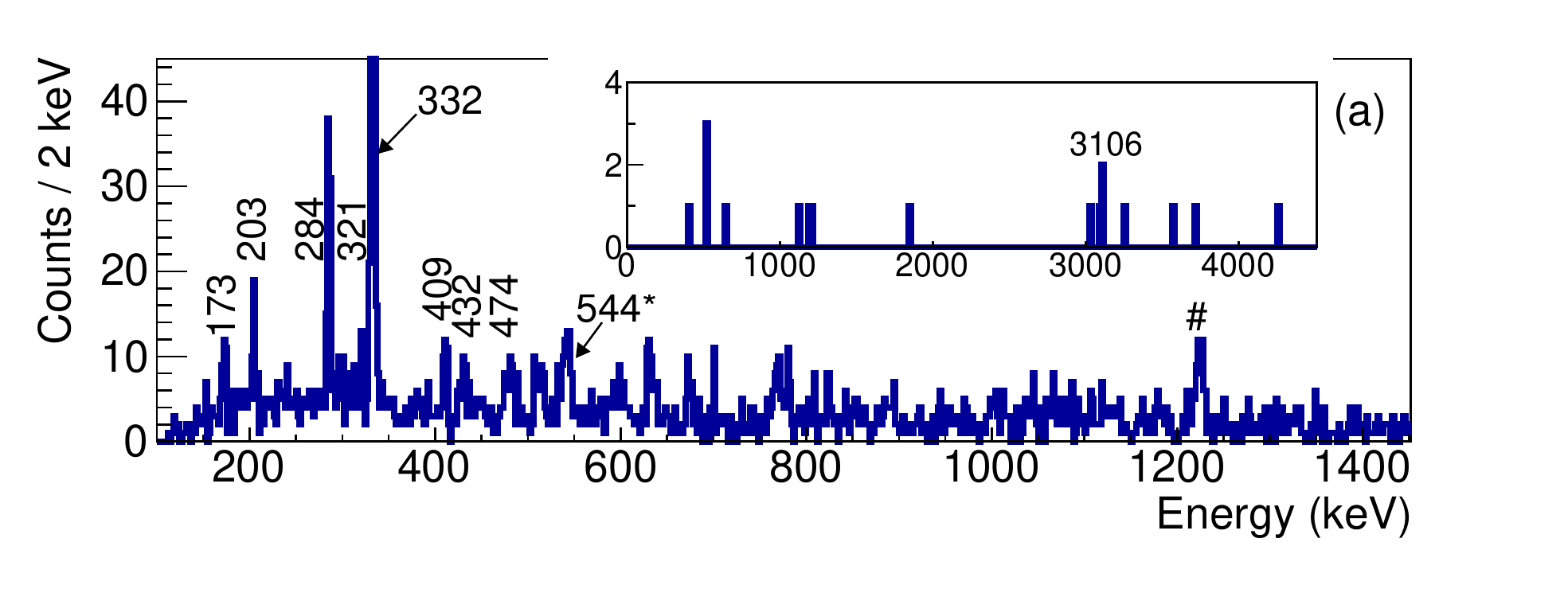}
    \includegraphics[trim=0cm 0cm 0cm 0.2cm, width=1.05\linewidth,clip]{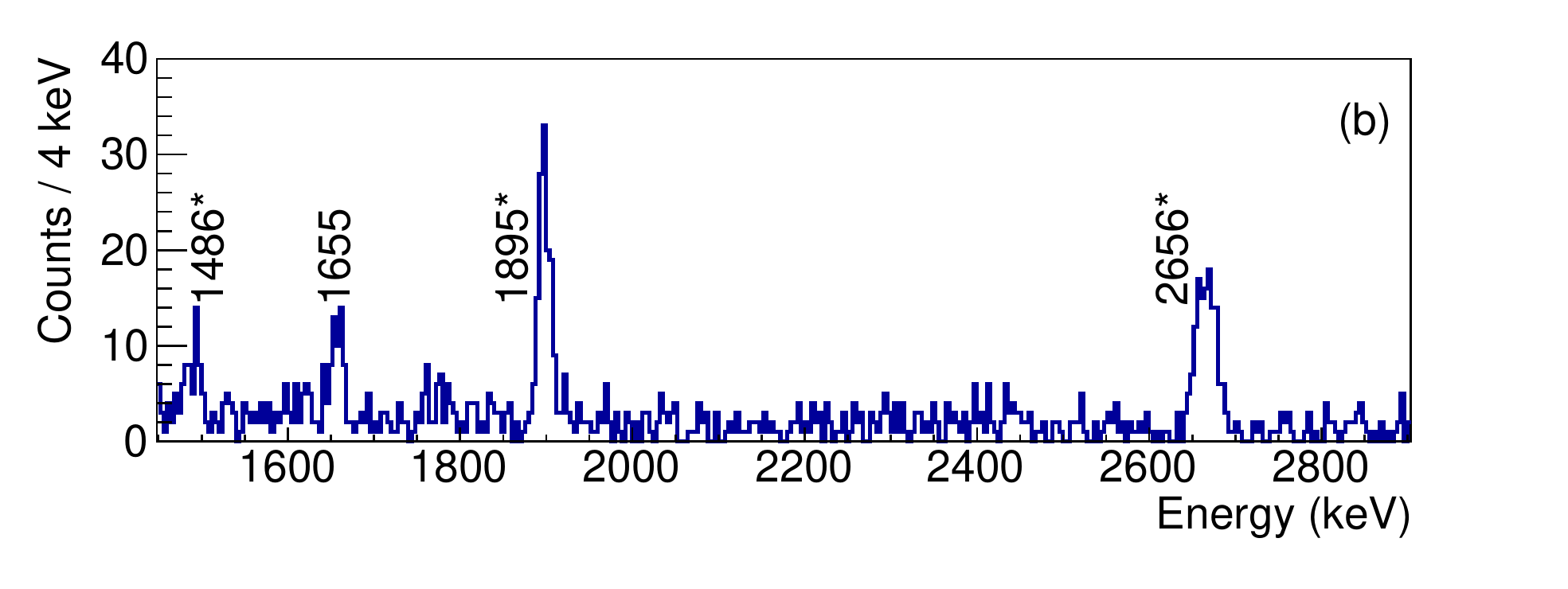}
     \includegraphics[trim=0cm 0cm 0cm 0.4cm, width=1.05\linewidth,clip]{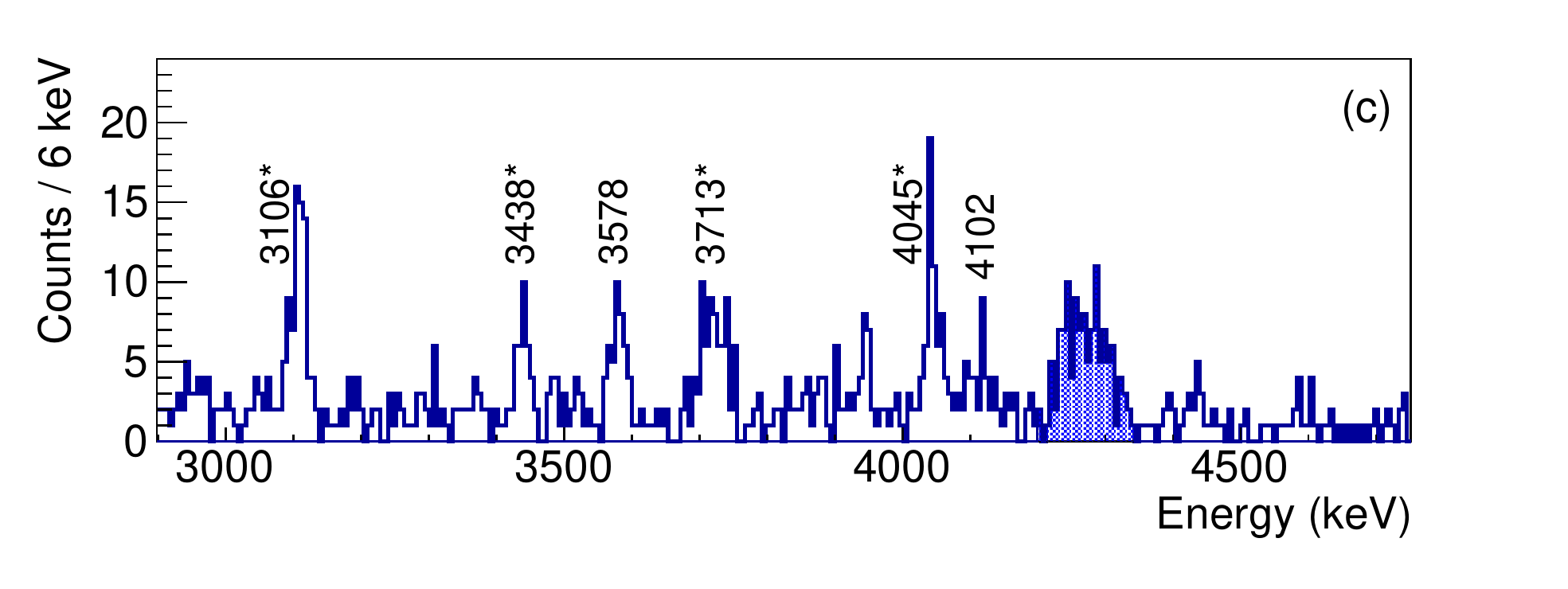}
        \caption{Spectrum of $\gamma$ rays following the $^{130}$Sn($^{9}$Be, $^{8}$Be $\gamma$)$^{131}$Sn reaction. The Doppler correction was made using the $\beta$ of the recoil optimized for the 332-keV transition in $^{131}$Sn for panel (a), and the 1895~keV transition for panels (b) and (c).  The $\gamma$ rays marked with a * are seen here for the first time.  The inset in panel (a) shows the coincidence spectrum (8~keV binning) gated on the 332-keV transition.  A known state in the contaminant $^{131}$Sb is marked  \#.  The shaded peak in panel (c) contains the known 4220-, 4247- and 4273-keV, and possibly other transitions to the 11/2$^-$ isomer that are unresolved here \cite{Bha01, Kha06} (color online).}
        \label{fig:spectrum}
        \end{figure}

\begin{figure}[ht]
 \includegraphics[trim=0cm 2cm 0cm 0.cm,width=1.05\linewidth,clip]{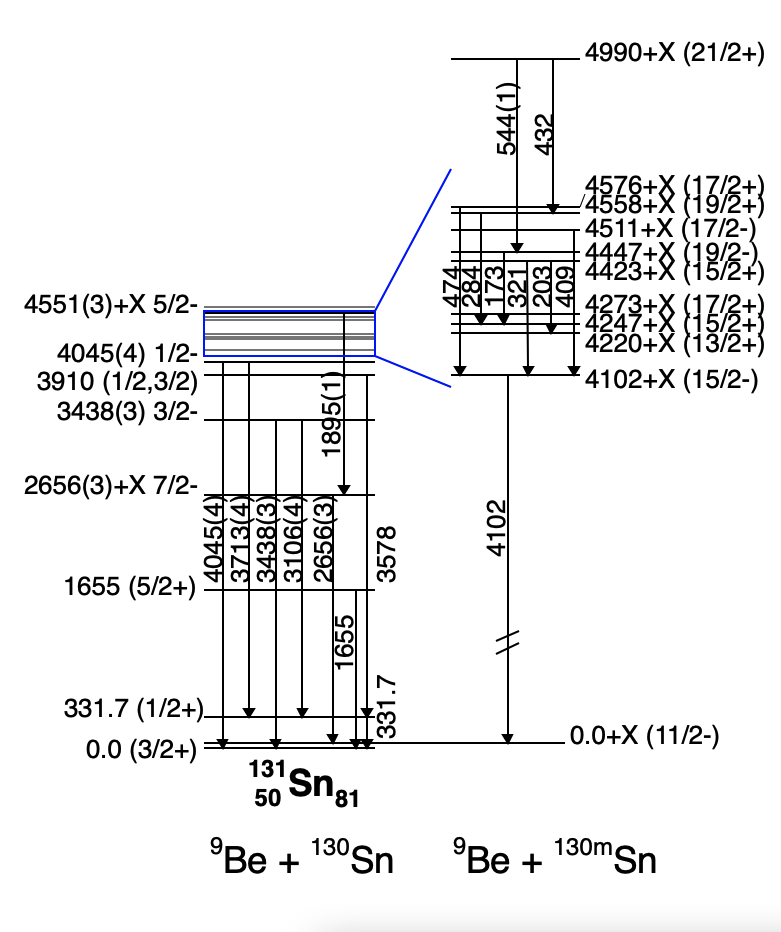}
\caption{Partial level scheme of $^{131}$Sn.  The $\gamma$ rays observed for the first time, and the states they de-excite, are shown with energy uncertainties in parentheses.  The known states \cite{Bha01,Kha06} with spins of 13/2 to 19/2 that were populated via the transfer on the 7$^-$ isomer in $^{130}$Sn are enclosed in the blue box and expanded to the right.  The excitation energy of the 11/2$^{-}$ isomeric state is currently unknown and therefore is labeled with an X (color online). }
        \label{fig:LevelScheme}
\end{figure}

The $\gamma$ transitions in Figure \ref{fig:spectrum} are shown in the level scheme, Figure \ref{fig:LevelScheme}.  Transitions that were observed for the first time are denoted by * in the spectrum and uncertainties in parentheses in the level scheme. The previously unobserved 510(3)-, 598(4)-, 632(2)-, and 1486(2)-keV transitions could not be placed in the level scheme. A known $\gamma$ ray in $^{131}$Sb following one-neutron transfer on the main contaminant in the beam is marked with a \#. The $\gamma$ rays for the (7/2$^-$), (3/2$^-$), (1/2$^-$), and (5/2$^-$) states are placed in the level scheme by comparing with the excitation energies from the $^{130}$Sn(d,p)$^{131}$Sn reaction reported in Kozub et al \cite{Koz12}.  

The 2656(3)-keV $\gamma$ ray has been assigned to the depopulation of the (7/2$^-$) state to the (11/2$^-$) isomer as it is the only strong transition that is close in energy to the known $\ell=3$ state.   Two $\gamma$ rays were observed from the (3/2$^-$) state, the 3438(3)-keV transition to the ground state and the 3106(4)-keV transition to the 331.7-keV state.   There were two decays observed from the (1/2$^-$) state at 4045(4)~keV.  For the 3438- and 4045-keV states the excitation energies deduced from the transitions to the (1/2$^+$) and ground state agree within uncertainties. A single 1895(1)-keV $\gamma$ transition from the (5/2$^-$) state to the 2656(3)+X-keV state is observed, defining the energy of that state above the isomer.
Table \ref{energies} compares the current and previously measured energies of the 1p-2h single-particle-like states.  The 3/2$^-$ and 1/2$^-$ states are the most important for neutron capture in the r-process and can be compared directly as they do not depend on the energy of the (11/2$^-$) isomer, marked as X in Figure \ref{fig:LevelScheme}.   The energies deduced for the (3/2$^-$) and (1/2$^-$) states in the present work are higher than those from the (d,p) measurement by around the 50~keV uncertainty quoted in that work, 34~keV and 59~keV, respectively.  

The neutron capture cross-section on $^{131}$Sn is expected to be dominated by the direct-semidirect (DSD) process, owing to the low level density.  DSD calculations were made using CUPIDO \cite{Par95, Arb05} in the way described in \cite{Man19} with the Koning-Delaroche potential \cite{Kon03} and fixed bound state parameters, $r_0 = 1.25$~fm and $a_0 = 0.65$~fm. The neutron capture cross-section calculation at the astrophysically important energy of E$_n$ = 30~keV, using the new excitation energies gives $\sigma = 115.9 \pm 20\mu$b, in agreement with previous results \cite{Man19, Moh12}.  The uncertainties are dominated by the spectroscopic factor.

The other two 1p-2h states decay to the low-lying isomer, whose energy is uncertain.  The energy of the (11/2$^-$) isomer was deduced by Fogelberg to be 65~keV from the placement of a $\gamma$ ray ``without firm support of a coincident $\gamma$" following the $\beta$ decay of $^{131}$In \cite{Fog04}.  The energy of the $\gamma$ ray was reproduced in a recent $\beta$-decay measurement \cite{Dun19}, again with no coincidence $\gamma$ rays to confirm its placement.  We are not able to improve the knowledge of the energy of this isomer from the current work.  To definitively find the energy of the (11/2$^-$) isomer would require finding a state that has a branch to the ground state and another to the isomer, or to perform a high precision mass measurement. Considering the coincidence rate seen in the inset of Figure \ref{fig:spectrum}(a) from the strong 331.7-keV transition, a much larger-statistics study would be needed to see such a branch from what would be a much weaker transition.  Using E$_x$~=~65~keV for the isomer, the (7/2$^-$) state is at E$_x$~=~2721~keV, 93~keV above the previous value, and the (5/2$^-$) state is at E$_x$~=~4616~keV, 39~keV lower than previously measured \cite{Koz12}.
\begin{table}
\caption{\label{energies}
Excitation energies and J$^{\pi}$ assignments of 1p-2h states in $^{131}$Sn from the current work and following the $^{130}$Sn(d,p)$^{131}$Sn reaction in Kozub et al.  \cite{Koz12}.  The J$^{\pi}$ assignments are confirmed here, as discussed in text.  }
\begin{tabular}{ccc} \toprule 
 \textrm Current & Kozub et al. \cite{Koz12} & \textrm {J$^{\pi}$} \\
  \textrm (keV)	& (keV)				  & \\
 \midrule 
2656(3) + X	  	& 2628(50)	& 7/2$^-$	\\

3438(3)  			& 3404(50)	& 3/2$^-$	\\

4045(4)			& 3986(50) 	& 1/2$^-$	\\

4551(3) + X 	 	& 4655(50)	&  5/2$^-$   \\  \bottomrule 
\end{tabular}
\end{table}
\subsection{Spin-parity assignments}
As spin-flip transitions are preferred in heavy-ion induced single-nucleon transfer reactions \cite{Bri72, Bon74}, the relative intensities of states populated in two carefully-chosen reactions can reveal the J$^{\pi}$ of the populated state, not just the $\ell$, as in an individual transfer reaction measurement.  This effect has been demonstrated in the comparison of the ($^{16}$O,$^{15}$N) and ($^{12}$C,$^{11}$B) reactions, in which the relative cross-sections for the population of known proton single-particle states were found to have a strong {\it j} dependence \cite{Kov72}.   The inverse-kinematics technique using RIBs has been validated using the ($^{9}$Be,$^{8}$Be) and ($^{13}$C,$^{12}$C) reactions \cite{All14, Rad02, All12}.  The current work is a continuation of those measurements, this time with a $^{130}$Sn beam.    As $^9$Be has an unpaired p$_{3/2}$ neutron there is a preference in the ($^{9}$Be,$^{8}$Be) reaction for populating states with $j_<=\ell - 1/2$.  However, this tendency is moderated by the weak binding of  $^9$Be (S$_n = 1.66$~MeV).  In contrast, ($^{13}$C,$^{12}$C) preferentially populates states with $j_>=\ell + 1/2$, owing to the dominantly p$_{1/2}$ nature of the last neutron in $^{13}$C.
\begin{figure}[ht]
       \includegraphics[trim=0.5cm 0.55cm 1cm 0.5cm,width=1. \linewidth, clip]{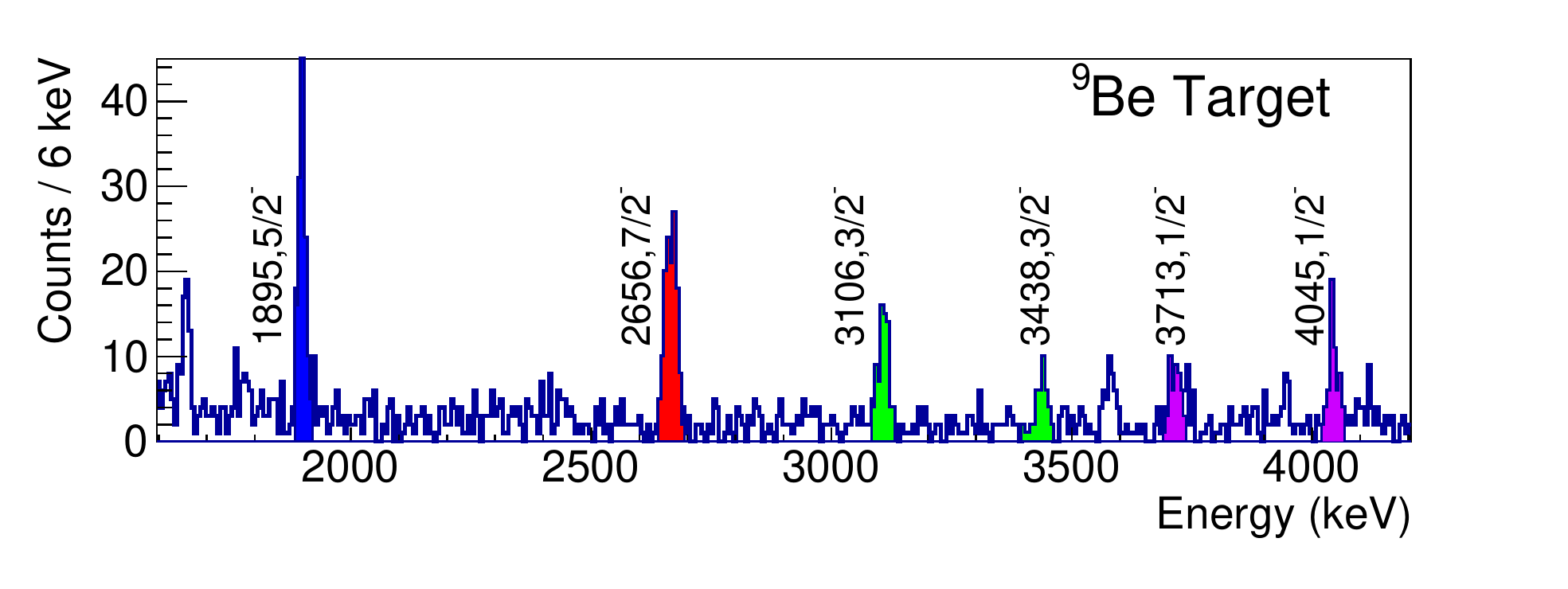} 
        \includegraphics[trim=0.5cm 0.55cm 1cm 0.5cm,width=1. \linewidth, clip]{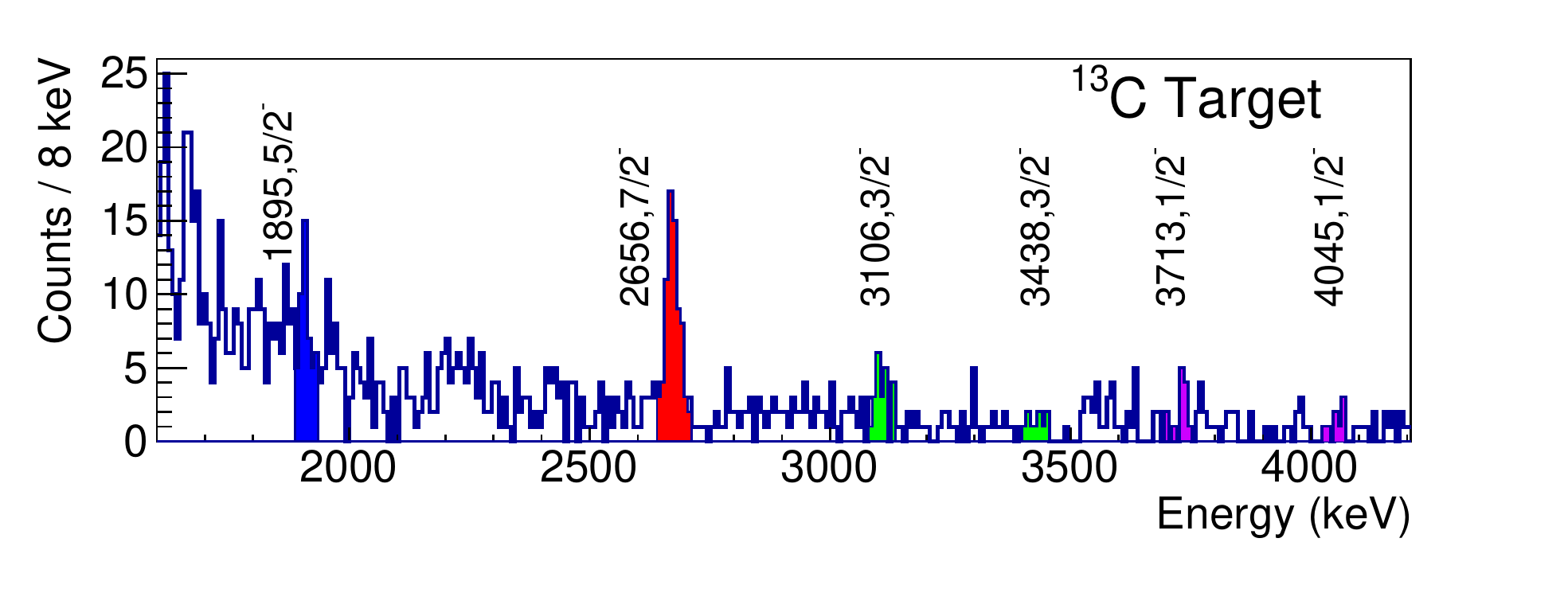}
         \caption{Comparison of the spectra of $\gamma$ rays emitted from states with the same $\ell$ value.  The $\gamma$ rays from the $\ell$ = 2 states at 4551+X keV and 2656+X keV are shown in blue and red, respectively.  The $\gamma$ rays from the $\ell$ = 1 states at 3438 keV and 4045 keV are shown in green and purple, respectively.  The efficiency-corrected intensities are given in Table \ref{ratios} (color online). }
        \label{fig:compare}
\end{figure}    
The relative cross sections for these states were measured using the $\gamma$-ray intensities (shown in Figure \ref{fig:compare}) normalized to the number of elastically scattered beam particles detected in the innermost ring of BareBall.  The figure does not show the optimal Doppler correction for all peaks, as one value of $\beta$ had to be selected for all the lines, resulting in some peaks being centered away from their true energy, and counts being dispersed across more channels.  A truer comparison can be made by comparing the total population of these states extracted from the $\gamma$ lines with the appropriate Doppler correction, background subtraction, and efficiency correction, accounting for feeding, and after normalizing to the elastic scattering.  The ratios of these populations from the $^9$Be and $^{13}$C targets are shown in Table \ref{ratios}. The 1895-keV $\gamma$ ray associated with the population of the 4551 + X~keV state was observed with almost four times the rate in the ($^9$Be,$^8$Be) compared to the ($^{13}$C,$^{12}$C) reaction.  This indicates that the 4551 + X~keV state has $j_<=\ell - 1/2$, confirming the assignment as 5/2$^-$ from Kozub et al \cite{Koz12}.  Conversely, the 2656~keV $\gamma$ ray is observed more strongly in the reaction on the $^{13}$C target, due to the $j_>=\ell + 1/2$ nature of the 7/2$^-$ state populated.  The two $\ell$~=~1 states are differentiated in the same way, although this is complicated by the poor statistics resulting from the low efficiency for very high energy $\gamma$ rays and weak population. It should be noted that the counts at 4045~keV from the $^{13}$C target were not considered to be above background; the coloring in Figure \ref{ratios} is to show the expected location for the peak.   The J$^{\pi}$ assignments of all four states discovered in \cite{Koz12} are confirmed.
\begin{table}
\caption{\label{ratios}
Ratios of normalized, efficiency-corrected intensities of $\gamma$ peaks produced in the $^{130}$Sn($^9$Be,$^8$Be)$^{131}$Sn and $^{130}$Sn($^{13}$C,$^{12}$C)$^{131}$Sn reactions, after background subtraction and accounting for feeding.  Since the reaction on the carbon target preferentially populates the j$_>~=\ell + 1/2$ states, a higher intensity ratio, compared to the partner state, indicates a j$_>$ assignment.  The numbers in parenthesis are statistical uncertainties.}
\begin{tabular}{ccccc}\toprule
\textrm{E$ _x $} & J$^{\pi}$ &  \multicolumn{2}{c}{\textrm{Normalized $\gamma$ Intensity}} & \textrm{Ratio} \\
\cmidrule(r){3-4} 
 \textrm  &  & $^9$Be target & $^{13}$C  target &  $^{13}$C/$^9$Be \\
 \midrule 
4551(3)+ X 	& 5/2$^-$ 	& 0.15(2)	&	0.04(1)	&	0.26(9)	\\
2656(3) + X	 & 7/2$^-$	& 0.07(1)	&	0.19(3)	&	2.8(6)	\\
4045(4) 		& 1/2$^-$ 	& 0.32(7)	&	$<$ 0.006	&	$<$ 0.017	\\
3438(3) 		& 3/2$^-$	&  0.13(2)	&	0.05(1)	&	0.4(1)	  \\ \bottomrule 
\end{tabular}
\end{table}

\subsection{Transfer on the 7$^-$ isomer}
One-neutron transfer at these energies typically transfers up to three units of angular momentum.  The ($^9$Be,$^8$Be) reaction on an even-even nucleus would therefore be expected to populate states in the final nucleus with J$\leq$ 7/2 .  The 173-, 203-, 284-, 321-, 409-, 432-, 474-, and 544-keV transitions seen in Figure \ref{fig:spectrum} are known $\gamma$ rays from states with between 15/2 and 21/2 units of spin \cite{Bha01,Kha06}, as shown on the right-hand side of Figure \ref{fig:LevelScheme}.  Additionally, the 4102-keV line is the known depopulation of a 15/2$^-$ state to the 11/2$^-$ isomer.  The population of these states cannot be explained through transfer on the 0$^+$ ground state of $^{130}$Sn. However, transferring up to three units of angular momentum on the 7$^-$ isomer provides enough angular momentum for the population of these high-spin states. The observation of these transitions provides direct evidence of transfer on the 7$^-$ isomer in the $^{130}$Sn beam, as shown on the right side of Figure \ref{fig:LevelScheme}. Transfer of p- and f-wave neutrons on the 7- isomer would lead to positive parity states being populated.  Transitions from negative-parity states were also seen, and can be explained as being populated through $\gamma$ decay from high-lying positive-parity states as can be seen for the 4447+X keV 19/2- state. 
A similar amount of the 7$^-$ isomer was most likely present during the $^{130}$Sn(d,p)$^{131}$Sn measurement, which would have produced a small level of contamination in the Q-value plot, but was otherwise unobservable.  

The comparison of the population of states from the measured $\gamma$-ray intensities depends on correctly taking into account the feeding from higher-lying levels.  Where those levels are observed, this feeding is taken into account.  However, it is possible that there is unobserved, indirect feeding. The $\gamma$-ray intensities in and out of the 332-keV state are shown in Table \ref{Intensities}, with 80\% of the intensity of the de-excitation accounted for through the feeding from the 3438-, 3910-, and 4045-keV states.  In contrast, only 40\% of the $\gamma$ intensity out of the 4102+X keV state, a tentative (15/2$^-$) state that is populated via an $\ell = 1$ transfer on the 7$^-$ isomer, is accounted for by feeding from above, albeit with significant uncertainties.\\
\begin{table}
\caption{\label{Intensities}
Intensity balance of the 332-keV 1/2$^+$ and 4102-keV+X (15/2$^-$) states.  The measured $\gamma$-ray normalized intensity into the 332-keV (4102-keV) states is the sum of the normalized intensities for the 3106-, 3578-, and 3713-keV (321-, 409- and 474-keV) states.  The fraction of the measured intensity out of the state that is explained by the measured intensity into the state is also given. }
\begin{tabular}{ccccc}\toprule 
 \textrm E$_x$ & E$_{\gamma}$ & \multicolumn{2}{c}{Normalized  Intensity} & Fraction\\
\cline{3-4}
 \textrm  (keV)& (keV)&  Out&  In & \\
 \midrule 
332	  	& 332 & 0.5(1)	&  0.39(8)	& 0.8 \\

3438  	& 3106 & 0.19(6)	& 	&\\

3910		&3578 & 0.019(9)	&	&\\

4045		& 3713 & 0.18(6)	& 	&\\

\midrule 

4102+X	& 4102	& 0.14(6)	&	0.06(1) & 0.4\\

4423+X	& 321 	& 0.019(6)	&		&	\\

4511+X	& 409	& 0.0034(8) &		&	\\

4576+X	& 474	&0.038(8)	&		& \\

 \bottomrule 
 \end{tabular}
\end{table}
There is not a straight-forward method to compare the population of individual states from transfer on the ground and isomeric states.  Table \ref{Intensities} suggests a large amount of unobserved feeding into the 4102+X-keV state, which has to be tempered by the large uncertainties coming from the low number of counts in the 4102-keV peak.  With those caveats, and assuming that the majority of the $\gamma$ intensity from the high-spin states decays through the 4102-keV state, a comparison can be made between the intensity of the $\gamma$ ray from the 4102+X-keV and the sum of the intensities of the $\gamma$ rays de-exciting the 1p2h states.  This gives an estimated 12\% of the cross-section coming from transfer on the isomer.  The isomeric component of the beam was not measured for this experiment, but is expected from previous measurements to be between 9 and 13\%. 

Ideally, nuclear structure conclusions would be drawn by making comparisons to modern calculations.  However, calculating accurate spectra for 1p-2h states across the N~=~82 shell closure is a challenge for modern nuclear structure theories.  There have been some recent successes with IMSGR calculations around $^{132}$Sn \cite{Miy21}, but these do not include spectra of these types of states.  Coupled-cluster calculations would require the inclusion of corrections that are beyond current capabilities for these heavier nuclei.  However, it is expected that in the next few years, one or both of these methods will be available for the states measured here and thus, these measurements will provide an essential benchmark. 
\section{Summary}
In summary, $\gamma$ rays from the de-excitation of states populated through the ($^9$Be,$^8$Be) and ($^{13}$C,$^{12}$C) reactions in inverse kinematics on a beam of $^{130}$Sn have allowed us to define the energies of 1p-2h states in $^{131}$Sn with greatly improved resolution.  The knowledge of the excitation energies of the 5/2$^-$ and 7/2$^-$ states is limited by the unknown energy of the 11/2$^-$ isomeric state to which they decay.  The relative intensities of $\gamma$ rays from the 4551-keV and 2656-keV states in the two reactions confirm their J$^{\pi}$ assignments as 5/2$^-$ and 7/2$^-$, respectively.  In a similar way, the 4045-keV and 3438-keV states have been confirmed as 1/2$^-$ and 3/2$^-$, respectively.

The population of a collection of higher-spin states above 4100 keV signals one-neutron transfer on the 7$^-$ isomeric state in $^{130}$Sn.   The population of these states corresponds approximately to the amount of isomer in the beam, suggesting that, as could be expected, transfer on the isomer is neither preferred, nor hindered, compared to transfer on the ground state. This is the first measurement of transfer on an isomer on a beam in this region of the chart of the nuclides. Modern nuclear structure theories will be able to compute the 1p-2h states in this region in the next few years, allowing meaningful comparison with the results here. These findings emphasize the need to include astromers in the $^{132}$Sn region in r-process nucleosynthesis simulations.
\begin{acknowledgements}

This work was supported by the US Department of Energy, Office of Science, Office of Nuclear Physics under contract number DE-SC001174 and DE-FG02-96ER40983 (UTK), No. DE-AC05-00OR22725 (ORNL), and DE-FG02-96ER40955 (TTU) and the National Science Foundation under Contract No. NSF- PHY-1067906  and PHY-1812316 (Rutgers). This research was sponsored in part by the National Nuclear Security Administration under the Stewardship Science Academic Alliances program under contract DE-FG52-08NA28552 and DE-NA0002132. This research was conducted at the Oak Ridge National Laboratory Holifield Radioactive Ion Beam Facility, a former D.O.E. Office of Science User Facility.The authors are grateful to the HRIBF facility operation staff efforts who made the measurements feasible.
\end{acknowledgements}

\bibliography{130Sn}

\end{document}